\documentclass[showpacs,preprintnumbers,superscriptaddress]{revtex4}
\usepackage{}

\usepackage{amsmath,amssymb,graphicx,bm}

\begin{document}

\title{Phase-space analysis of a class of k-essence cosmology}

\author{Rong-Jia Yang}
\email{yangrj08@gmail.com}
 \affiliation{College of Physical Science and Technology, Hebei University, Baoding 071002, China}

\author{Xiang-Ting Gao}
 \affiliation{College of Physical Science and Technology, Hebei University, Baoding 071002, China}

\date{\today}

\begin{abstract}
We perform a detailed phase-space analysis of a class of k-essence cosmology. We find the critical points can be divided into three classes: points unstable but the model stable, both points and the model stable, points stable but the model unstable. Like the case of points unstable but the model stable, the case of points stable but the model unstable is not relevant from a cosmological point of view, though they can be late-time attractors for the universe. So in order to study the possible final state of the universe, it is important to investigate not only the stability of the critical points but also the stability of the model. The case of both points and the model stable can further be divided into two classes: points only presenting decelerating phases and points at which all decelerating, constant-speed, and accelerating phases can appear; the final state of the universe dependents on the potential.


\end{abstract}

\maketitle

\section{Introduction}
In the last decade a convergence of independent cosmological
observations suggested that the Universe is experiencing accelerated
expansion. An unknown energy component, dubbed as dark energy, is
proposed to explain this acceleration. Dark energy almost equally
distributes in the Universe, and its pressure is negative. The simplest and most theoretically appealing candidate of
dark energy is the vacuum energy (or the cosmological constant
$\Lambda$) with a constant equation of state (EoS) parameter $w=-1$.
This scenario is in general agreement with the current astronomical
observations, but has difficulties to reconcile the small
observational value of dark energy density with estimates from
quantum field theories; this is the cosmological constant problem \cite{weinberg,Copeland2006}.
Recently it was shown that $\Lambda$CDM model may also suffer an age problem \cite{Yang2010}.
It is thus natural to pursue alternative possibilities to explain
the mystery of dark energy. Over the past decade numerous dark energy models have been proposed,
such as quintessence, phantom, k-essence, tachyon, (generalized) Chaplygin gas, DGP, quintom \cite{Feng}, etc. k-essence, a simple approach toward constructing a model for an accelerated expansion of the Universe, is to work with the idea that the unknown dark energy component is due exclusively
to a minimally coupled scalar field $\phi$ with non-canonical kinetic energy which results in the negative pressure \cite{Arm00}.

K-essence scenario has received much attention, it was originally proposed as a model for inflation
\cite{Damour1999}, and then as a model for dark energy \cite{Arm00}.
In several cases, k-essence cannot be observationally distinguished from quintessence \cite{Malquarti2003}.
A method to obtain a phantom version of FRW k-essence cosmologies was devised in \cite{Aguirregabiria2004}.
The stability of k-essence was studied in \cite{Abramo2006}.
Dynamics of k-essence were discussed in \cite{Rendall2006}.
Conditions for stable tracker solutions for k-essence in a general cosmological
background were derived in \cite{Das2006}.
Slow-roll conditions for thawing k-essence were obtained in \cite{Chiba2009}.
A connection between the holographic dark energy density and the kinetic k-essence energy density
was discussed in \cite{Cruz2009}.
An holographic k-essence model of dark energy was proposed in \cite{Granda2009}.
The geometrical diagnostic for purely kinetic k-essence dark energy was discussed in \cite{Gao2010}.
The equivalence of a barotropic perfect fluid with a k-essence scalar field was considered in \cite{Arroja2010,Feinstein2005}.
Dark matter to dark energy transition in k-essence cosmologies was examined in \cite{Chimento2009}.
Models of dark energy with purely kinetic multiple k-essence sources that allow for the crossing
of the phantom divide line were investigated in \cite{Sur2009, Chimento2006}.
The thermodynamic properties of of k-essence was discussed in \cite{Bilic2008}.
Models of k-essence unified dark matter were discussed in \cite{Daniele2007,Scherrer2004,Bose2009,Feinstein2006,Chim04}.
Power-law expansion in k-essence cosmology was investigated in \cite{Feinstein2004}.
Theoretical and observational Constraints on k-essence dark energy models were discussed in \cite{Yang2009,Yang2008,Yang2008a}.
In Ref. \cite{Sen2006}, a model independent method of reconstructing the Lagrangian for the k-essence field by using three parametrizations
for the Hubble parameter $H(z)$ was studied in detail. With assumptions on the EoS of k-essence as functions of the scale factor $a$, Ref. \cite{Putter2007}
discussed the forms of the Lagrangians. By assuming the EoS of k-essence as functions of the kinetic energy, Ref. \cite{yang10} considered restrictions on purely kinetic k-essence. Ermakov invariant of k-essence and its observational probes for the early universe are discussed in \cite{Gangopadhyay}.

Since from it was proposed, k-essence was been studied intensively. It is still worth
investigating in a systematic way the possible cosmological
behavior of the k-essence cosmology. Thus, in
the present work we perform a phase-space and stability analysis
of a class of k-essence cosmology. We are interesting in
investigating the possible late-time solutions. In these solutions
we calculate various observable quantities, such are the density of the
dark energy and the EoS parameters. As we see,
indeed k-essence cosmology can be consistent with
observations.

This paper is organized as follows, in the following section, we review the model of k-essence. In Sec. III, we perform a phase-space and stability analysis
of k-essence cosmology. Finally, we shall close with a few concluding remarks in Sec. IV.

\section{K-essence cosmology}
Scalar fields with non-canonical kinetic terms often appear in particle physics. In general the action for such theories can be expressed
as \cite{Arm00, Damour1999, Garriga1999}
\begin{eqnarray}
\label{action}
S=\int d^4x\sqrt{-g}\left[\frac{R}{2}+p(\phi,X)\right],
\end{eqnarray}
where $p(\phi,X)$ is a function in terms of a scalar field $\phi$ and its kinetic energy $X\equiv-\frac{1}{2}\partial_{\mu}\phi\partial^{\mu}\phi$.
We assume a flat and homogeneous Friedmann-Robertson-Walker (FRW) space-time and work in units $8\pi G=c=1$. The action (\ref{action}) includes
a wide variety of theories, such as low energy effective string theory \cite{Damour1999}, ghost condensate \cite{Cheng}, tachyon \cite{Padmanabhan, Keresztes}, and DBI theory \cite{Alishahiha}. Here
we consider an action which was extensively studied in literatures \cite{Damour1999,Scherrer2004, Gasperini2003, Bean2010,Piazza,Chiba2000}
\begin{eqnarray}
\label{p}
p_\phi=V(\phi)\left[-X+X^2\right]\equiv V(\phi)F(X).
\end{eqnarray}
where $V(\phi)$ is the k-essence potential. For a constant potential $V_0$, action (\ref{p}) are discussed in \cite{Scherrer2004,Chiba2000}. The energy density of k-essence take the forms
\begin{eqnarray}
\rho_\phi &=&V(\phi)[2XF_{X}-F]=V(3X^2-X),
\end{eqnarray}
here $F_{X}\equiv dF/dX$. The corresponding EoS parameter and the effective sound speed are
given by
\begin{eqnarray}
\label{w}w_\phi&=&\frac{F}{2XF_{X}-F}=\frac{X^2-X}{3X^2-X}, \\
\label{c}c^{2}_{\rm s}&=&\frac{\partial p/\partial
X}{\partial\rho /\partial X}=\frac{F_{X}}{F_{X}+2XF_{XX}}=\frac{2X-1}{6X-1},
\end{eqnarray}
with $F_{XX}\equiv d^{2}F/dX^{2}$. The definition of the sound speed
comes from the equation describing the evolution of linear
adiabatic perturbations in a k-essence dominated Universe \cite{Garriga1999} (the non-adiabatic perturbation was discussed in \cite{Unnikrishnan2010,Christopherson}, here we only consider the case of adiabatic perturbations). Perturbations can become classic unstable if the sound speed is imaginary, $c_{\rm s}^2<0$, so we insist on $c_{\rm s}^2 \geq 0$. Another potentially interesting requirement to consider is $c_{\rm s}^2 \leq 1$, which says that the sound speed should not exceed the speed of light, which suggests violation of causality. Note, however, this is still an open problem (see e. g. \cite{Babichev2008,Bruneton2007,Kang2007,Bonvin2006,Gorini2008,Ellis2007}).

In a flat and homogeneous FRW space-time, the equation for the k-essence field takes the form
\begin{eqnarray}
\label{field}
(F_{X}+2XF_{XX})\ddot{\phi}+3HF_{X}\dot{\phi}+(2XF_{X}-F)\frac{V_{\phi}}{V}=0,
\end{eqnarray}
where $V_{\phi}\equiv dV/d\phi$, $H$ is the Hubble parameter related to the Friedmann equations

\begin{eqnarray}
\label{f1}
H^2=\frac{1}{3}(\rho_{\rm m}+\rho_{\phi}),\\
\label{f2}
\dot{H}=-\frac{1}{2}(\rho_{\rm m}+\rho_{\phi}+p_{\phi}).
\end{eqnarray}
Although we could straightforwardly include baryonic matter and radiation in the model, we neglect them for simplicity. In next section, we will transform Eqs. (\ref{f1}) and (\ref{f2}) into an autonomous dynamical system to perform the phase-space and stability analysis.

\section{Phase-space analysis}
In general, in order to perform the phase-space and stability analysis, we have to transform the cosmological equations
into an autonomous dynamical system by introducing the auxiliary variable $x$ and $y$ \cite{Hirsch, Leon2009, Copeland1998, Ferreira1997, Copeland2006, Tsujikawa2008, Chen, Maeda, Aguirregabiria, Amendola, Guo, Guo1}. Using the auxiliary variables $x$ and $y$, the cosmological equations of motion, give a self-autonomous system $\textbf{X}'=\textbf{f}(\textbf{X})$, where $\textbf{X}$ is the column vector constituted by the auxiliary variables, $\textbf{f}(\textbf{X})$ the corresponding column vector of the autonomous equations, and the prime denotes a derivative with respect to the logarithm of the scale factor, $N\equiv \ln a$. The critical points $\textbf{X}_{\rm c}$ are extracted satisfying $\textbf{X}'=0$, and in order to determine the stability properties of these critical points we expand around $\textbf{X}_{\rm c}$, setting $\textbf{X}=\textbf{X}_{\rm c}+\textbf{U}$ with $\textbf{U}$ the perturbations of the variables considered as column vector. Thus, up to the first order we acquire $\textbf{U}'=\textbf{M}\cdot \textbf{U}$, where the matrix $\textbf{M}$ contains the coefficients of the perturbation equations. Thus, for each critical point, the eigenvalues of $M$ determine its type and stability. The condition for the stability of the critical points is: Tr $\textbf{M}<0$ and $\det \textbf{M}>0$.

\subsection{The basic equations and the critical points}
\begin{figure}
\includegraphics[width=10cm]{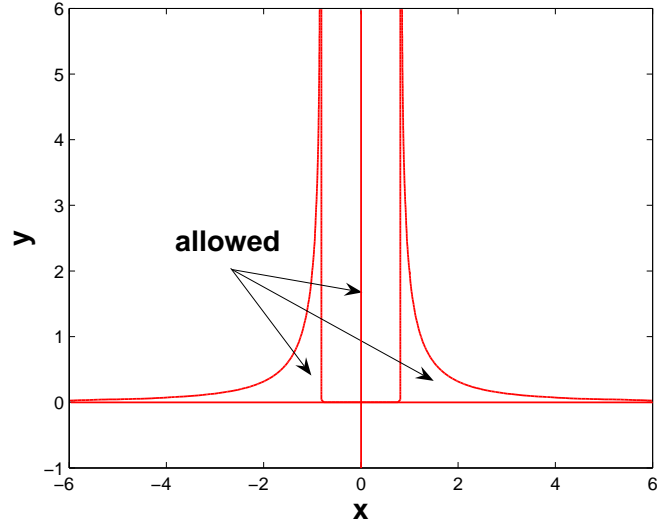}
\caption{The allowed regions of the auxiliary variable $x$ and $y$. \label{Fig1}}
\end{figure}

In order to transform the cosmological equations into an autonomous dynamical system, we have to introduce the auxiliary variable:
\begin{eqnarray}
x=\dot{\phi},~~~~y=\frac{\sqrt{V(\phi)}}{\sqrt{3}H}.
\end{eqnarray}
together with $N=\ln a$. It is easy to have $\frac{d}{dt}=H\frac{d}{dM}$. Using these variables we can straightforwardly obtain the density parameters of dark energy as
\begin{eqnarray}
\Omega_\phi=\frac{\rho_\phi}{3H^2}=\frac{1}{4}x^2y^2(3x^2-2).
\end{eqnarray}
Because $0\leq \Omega_\phi \leq 1$, we get the constraint equation
\begin{eqnarray}
0\leq \frac{1}{4}x^2y^2(3x^2-2)\leq 1.
\end{eqnarray}
in which the auxiliary variable $x$ and $y$ are allowed, see Fig. (\ref{Fig1}). In general, when the auxiliary variable $x$ and $y$ can take infinite values,
it is necessary to analyze the dynamics at infinity by using Poincar\'{e} Projection method \cite{Carloni, Carloni2006}. However, in the case we consider here, it is obviously
that when $|x|>4$, $y$ rapidly runs close to the zero; when $y>4$, $x$ rapidly runs close to $\sqrt{2/3}$; When $x=0$, $y$ take arbitrary values. So it is not necessary to analyze the
dynamics at infinity in the case we discussed.

The EoS, the sound speed of k-essence, and the total EoS are, respectively
\begin{eqnarray}
w_\phi&=&\frac{x^4-2x^2}{3x^4-2x^2},\\
\label{sound}
c^{2}_{\rm s}&=&\frac{x^2-1}{3x^2-1},\\
w_{\rm t}&=&\frac{p_\phi}{\rho_\phi+\rho_{\rm m}}=\frac{1}{4}x^2y^2(x^2-2),
\end{eqnarray}

A set of Eqs. (\ref{field}), (\ref{f1}) and (\ref{f2}) gives a self-autonomous system in terms of the auxiliary variable $x$ and $y$:
\begin{eqnarray}
\label{aut}
x' &=& -\frac{3x(x^2-1)}{3x^2-1}+\frac{\sqrt{3}}{4}\frac{\lambda yx^2(3x^2-2)}{3x^2-1},\\ \nonumber
y' &=& -\frac{\sqrt{3}}{2}\lambda xy^2+\frac{3}{2}y\left[1+\frac{1}{4}x^2y^2(x^2-2)\right].
\end{eqnarray}
where the prime denotes a derivative with respect to the logarithm of the scale factor, $N\equiv \ln a$, and $\lambda \equiv -V_\phi/V^{\frac{3}{2}}$. Here we only consider the case in which the $\lambda$ is a constant, that is to say $V(\phi)\propto \phi^{-2}$. Eq. (15)
form an autonomous dynamical system in this case, this self-autonomous system are valid in the whole phase-space, not only at the critical points.

The critical points $(x_{\rm c},y_{\rm c})$ of the autonomous system
(\ref{aut}) are obtained by setting the left hand sides of
the equations to zero, namely let $x'=y'=0$. After some algebraic calculus, we obtain the critical points as following:

Critical point $P_1$
\begin{eqnarray}
x_1 &=& 0,\\ \nonumber
y_1 &=& 0.
\end{eqnarray}
We calculate the
values of $c^{2}_{\rm s}$, $\Omega_{\phi}$, $w_{\phi}$, and $w_{\rm t}$, as shown in table \ref{crit}. See for example: $\Omega_{\phi}=0$, the universe is completely
dominated by dark matter. We also note that in this case, $w_{\phi}$ remains unspecified
and the results hold independently of its value.

Critical point $P_2$
\begin{eqnarray}
x_2 &=& 1,\\ \nonumber
y_2 &=& 0.
\end{eqnarray}
The corresponding cosmological parameters are presented in table \ref{crit}. Though in this case the k-essence behaves like the cosmological constant ($w_{\phi}=-1$),  the universe is completely dominated by dark matter ($\Omega_{\phi}=0$).

Critical point $P_3$
\begin{eqnarray}
x_3 &=& -1,\\ \nonumber
y_3 &=&0.
\end{eqnarray}
In this case, the behavior of the universe is like that at point $P_2$. Points $P_1$, $P_2$, and $P_3$ can be obtained in other scalar dark energy models (see e. g. \cite{Copeland2006}), but the corresponding cosmological parameters, such as $c^{2}_{\rm s}$, $\Omega_{\phi}$, $w_{\phi}$, and $w_{\rm t}$,
may be different because they are model-dependent.

Critical point $P_4$
\begin{eqnarray}
x_4 &=& -\sqrt{2},\\ \nonumber
y_4 &=& -\frac{\sqrt{6}}{2}\frac{1}{\lambda},
\end{eqnarray}
with $w_{\phi}=0$, meaning k-essence behaves like dark matter; $\Omega_{\phi}=3/\lambda^2$ ($\lambda<-\sqrt{3}$), a universe is partly
dominated by k-essence.
\begin{figure}
\includegraphics[width=10cm]{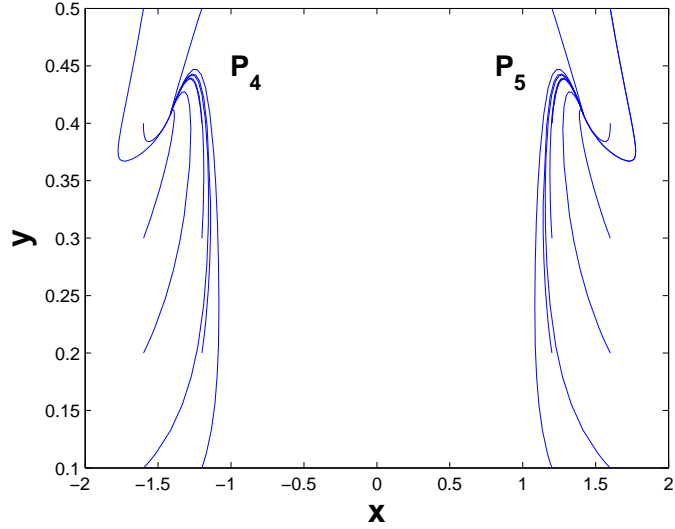}
\caption{Phase-space for k-essence cosmology, with the choice $\lambda=-3$ for critical point $P_4$ and $\lambda=3$ for critical point $P_5$. \label{Fig2}}
\end{figure}

Critical point $P_5$
\begin{eqnarray}
x_5 &=& \sqrt{2},\\ \nonumber
y_5 &=& \frac{\sqrt{6}}{2}\frac{1}{\lambda}.
\end{eqnarray}
In this case, the behavior of the universe for $\lambda> \sqrt{3}$ is like that at point $P_4$. We plot critical points $P_4$ for $\lambda=-3$ and $P_5$ for $\lambda=3$ in Fig. \ref{Fig2}.

Critical point $P_6$
\begin{eqnarray}
x_6 &=& -\sqrt{1+\frac{\lambda^2}{8}+\frac{\sqrt{48\lambda^2+9\lambda^4}}{24}},\\ \nonumber
y_6 &=& \frac{x_6(-20\sqrt{3}\lambda^2-3\sqrt{3}\lambda^4+12\sqrt{16\lambda^2+3\lambda^4}+3\lambda^2\sqrt{16\lambda^2+3\lambda^4})}{4\lambda(6+\lambda^2)}.
\end{eqnarray}
with $\Omega_{\phi}=1$, meaning a k-essence dominated universe; and $w_{\rm t}=-1-\frac{1}{2}\lambda^2+\frac{\sqrt{48\lambda^2+9\lambda^4}}{6}$ ($-\sqrt{3}<\lambda<0$), all decelerating, constant-speed, and accelerating phases can appear.

\begin{figure}
\includegraphics[width=10cm]{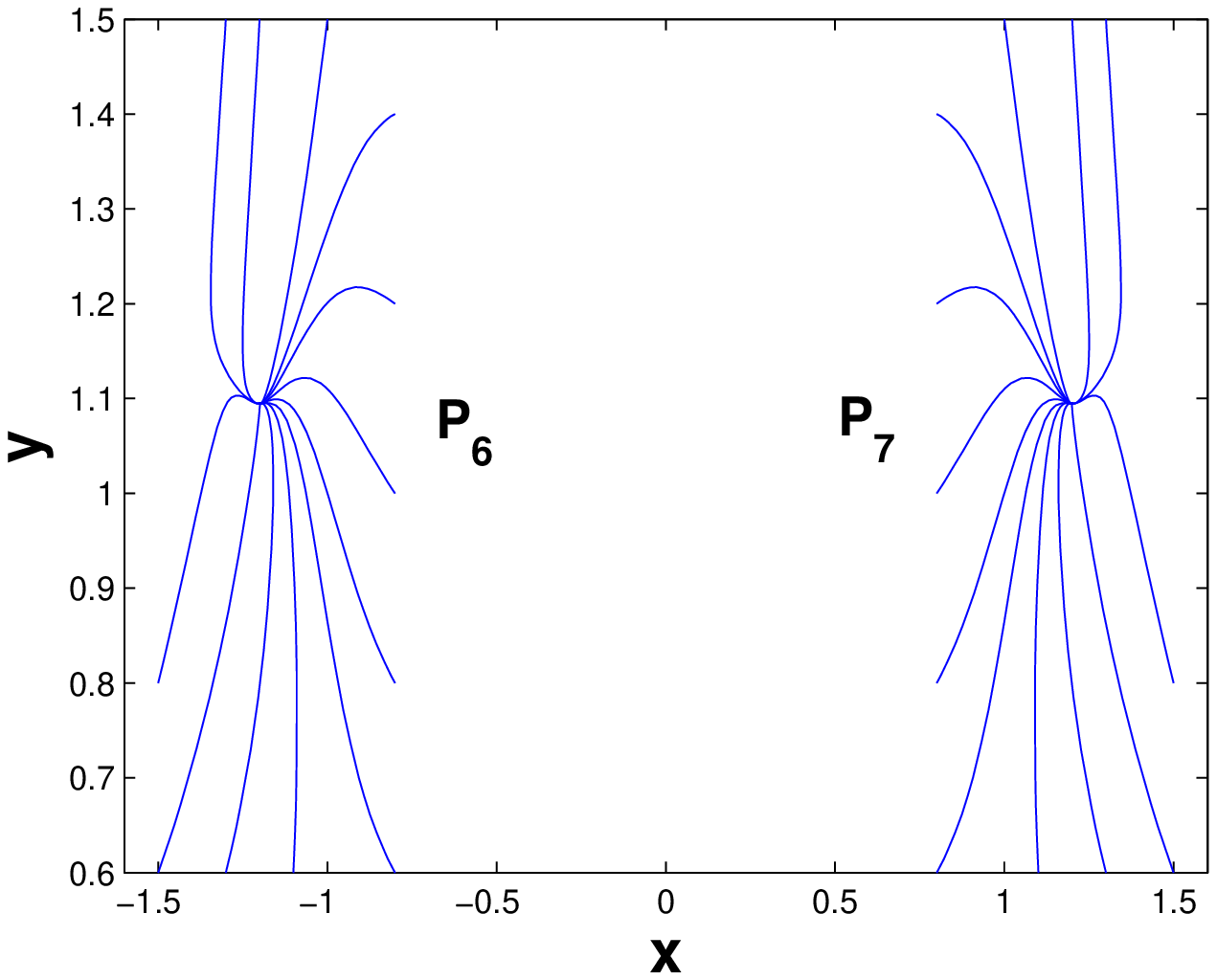}
\caption{Phase-space for k-essence cosmology, with the choice $\lambda=-1$ for critical point $P_6$ and $\lambda=1$ for critical point $P_7$. \label{Fig3}}
\end{figure}

Critical point $P_7$
\begin{eqnarray}
x_7 &=& \sqrt{1+\frac{\lambda^2}{8}+\frac{\sqrt{48\lambda^2+9\lambda^4}}{24}},\\ \nonumber
y_7 &=& \frac{x_7(-20\sqrt{3}\lambda^2-3\sqrt{3}\lambda^4+12\sqrt{16\lambda^2+3\lambda^4}+3\lambda^2\sqrt{16\lambda^2+3\lambda^4})}{4\lambda(6+\lambda^2)}
\end{eqnarray}
In this case, the behavior of the universe for $0<\lambda<\sqrt{3}$ is like that at point $P_6$.
We plot critical points $P_6$ for $\lambda=-1$ and $P_7$ for $\lambda=1$ in Fig. \ref{Fig3}.

Critical point $P_8$
\begin{eqnarray}
x_8 &=& -\frac{\sqrt{144+18\lambda^2-3\sqrt{32\lambda^2+6\lambda^4}}}{12},\\ \nonumber
y_8 &=& \frac{x_8(-20\sqrt{3}\lambda^2-3\sqrt{3}\lambda^4-12\sqrt{16\lambda^2+3\lambda^4}-3\lambda^2\sqrt{16\lambda^2+3\lambda^4})}{4\lambda(6+\lambda^2)}
\end{eqnarray}
with $\Omega_{\phi}=1$, a k-essence dominated universe; $w_{\rm t}=-1-\frac{1}{2}\lambda^2-\frac{\sqrt{48\lambda^2+9\lambda^4}}{6}$ ($\lambda>0$), only accelerating phases can appear.

\begin{figure}
\includegraphics[width=10cm]{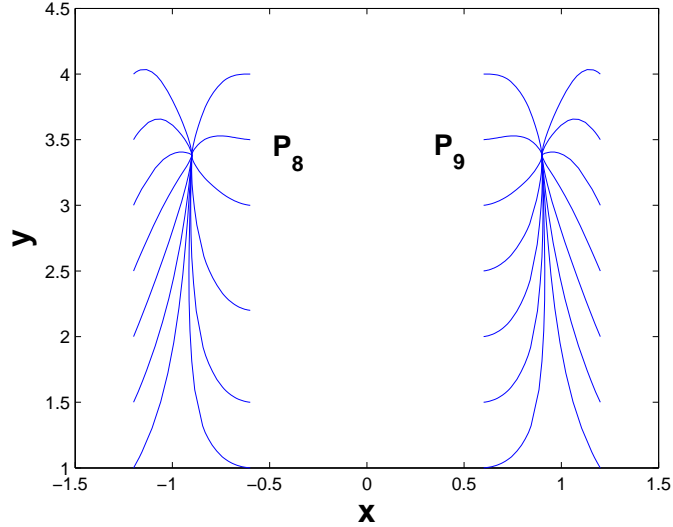}
\caption{Phase-space for k-essence cosmology, with the choice $\lambda=-1$ for critical point $P_8$ and $\lambda=1$ for critical point $P_9$. \label{Fig4}}
\end{figure}

Critical point $P_9$
\begin{eqnarray}
x_9 &=& \frac{\sqrt{144+18\lambda^2-3\sqrt{32\lambda^2+6\lambda^4}}}{12},\\ \nonumber
y_9 &=& \frac{x_9(-20\sqrt{3}\lambda^2-3\sqrt{3}\lambda^4-12\sqrt{16\lambda^2+3\lambda^4}-3\lambda^2\sqrt{16\lambda^2+3\lambda^4})}{4\lambda(6+\lambda^2)}
\end{eqnarray}
In this case, the behavior of the universe for $\lambda<0$ is like that at point $P_8$. We plot critical points $P_8$ for $\lambda=1$ and $P_9$ for $\lambda=-1$ in Fig. \ref{Fig4}.

\subsection{Stability analysis and the attractors}
Now we analyze stability of the critical points obtained above. For hyperbolic critical
points (all the eigenvalues have real parts different from zero)
one can easily extract their type (source (unstable) for positive
real parts, saddle for real parts of different sign and sink
(stable) for negative real parts). However, if at least one
eigenvalue has a zero real part (non-hyperbolic critical point)
one is not able to obtain conclusive information about the
stability from linearization and needs to resort to other tools
like Normal Forms calculations  \cite{arrowsmith,wiggins}, or
numerical experimentation. The $2\times2$ matrix ${\bf {M}}$ of the linearized perturbation
equations writes:
\[\textbf{M}= \left[ \begin{array}{ccc}
 -3+\sqrt{3}\lambda x_{\rm c}y_{\rm c} & & \frac{\sqrt{3}}{4}\frac{\lambda x^2_{\rm c}(3x^2_{\rm c}-2)}{3 x^2_{\rm c}-1} \\
 -\frac{\sqrt{3}}{2}\lambda y^2_{\rm c}+\frac{3}{2}x^3_{\rm c}y^3_{\rm c}-\frac{3}{2}x_{\rm c}y^3_{\rm c} & & -\sqrt{3}\lambda x_{\rm c}y_{\rm c}-\frac{9}{4} x^2_{\rm c}y^2_{\rm c}+\frac{9}{8} x^4_{\rm c}y^2_{\rm c}+\frac{3}{2}
\end{array} \right],\]
According to the conditions for the stability of the critical points: Tr $\textbf{M}<0$ and $\det \textbf{M}>0$, we obtain the ranges of $\lambda$
to make the critical points obtained above stable, as shown in table \ref{crit} in which we also present the necessary conditions for their existence,
as well as the corresponding cosmological parameters, $c^{2}_{\rm s}$, $\Omega_{\phi}$, $w_{\phi}$, and $w_{\rm t}$, with these parameters we can determine
the stability of the model and discuss whether there exists acceleration phase or not.
\begin{table*}
\begin{center}
\begin{tabular}{|c|c|c|c|c|c|c|c|}
  \hline
  Cr. P & Existence & Stable for & $c^{2}_{\rm s}$ & $\Omega_{\phi}$ & $w_{\phi}$ & $w_{\rm t}$ & Acceleration \\\hline
  $P_1$ & $\lambda=0$ & None & 1 & 0 & arbitrary & 0 & No \\\hline
  $P_2$ & $\lambda=0$ & None & 0 & 0 & -1 & 0 & No \\\hline
  $P_3$ & $\lambda=0$ & None & 0 & 0 & -1 & 0 & No \\\hline
  $P_4$ & $\lambda \leq -\sqrt{3}$ & $\lambda<-\sqrt{3}$ & $\frac{1}{5}$ & $\frac{3}{\lambda^2}$ & 0 & 0 & No \\\hline
  $P_5$ & $\lambda \geq \sqrt{3}$ & $\lambda > \sqrt{3}$ & $\frac{1}{5}$ & $\frac{3}{\lambda^2}$ & 0 & 0 & No \\\hline
  $P_6$ & $\lambda <0$ & $-\sqrt{3}<\lambda<0$ & $\frac{3\lambda^2+\sqrt{48\lambda^2+9\lambda^4}}{48+9\lambda^2+3\sqrt{48\lambda^2+9\lambda^4}}$ & 1 & $\frac{-1+\frac{\lambda^2}{8}+\frac{\sqrt{48\lambda^2+9\lambda^4}}{24}}{1+\frac{3\lambda^2}{8}+\frac{\sqrt{48\lambda^2+9\lambda^4}}{8}}$ & $-1-\frac{1}{2}\lambda^2+\frac{\sqrt{48\lambda^2+9\lambda^4}}{6}$ & $-\frac{\sqrt{6}}{3}< \lambda <0$ \\\hline
  $P_7$ & $\lambda >0$ & $0<\lambda<\sqrt{3}$ & $\frac{3\lambda^2+\sqrt{48\lambda^2+9\lambda^4}}{48+9\lambda^2+3\sqrt{48\lambda^2+9\lambda^4}}$ & 1 & $\frac{-1+\frac{\lambda^2}{8}+\frac{\sqrt{48\lambda^2+9\lambda^4}}{24}}{1+\frac{3\lambda^2}{8}+\frac{\sqrt{48\lambda^2+9\lambda^4}}{8}}$ & $-1-\frac{1}{2}\lambda^2+\frac{\sqrt{48\lambda^2+9\lambda^4}}{6}$ & $0< \lambda < \frac{\sqrt{6}}{3}$ \\\hline
  $P_8$ & $\lambda >0$ & $\lambda >0$ & $\frac{3\lambda^2-\sqrt{48\lambda^2+9\lambda^4}}{48+9\lambda^2-3\sqrt{48\lambda^2+9\lambda^4}}$ & 1 & $\frac{-1+\frac{\lambda^2}{8}-\frac{\sqrt{48\lambda^2+9\lambda^4}}{24}}{1+\frac{3\lambda^2}{8}-\frac{\sqrt{48\lambda^2+9\lambda^4}}{8}}$ & $-1-\frac{1}{2}\lambda^2-\frac{\sqrt{48\lambda^2+9\lambda^4}}{6}$ & $\lambda >0$ \\\hline
  $P_9$ & $\lambda <0$ & $\lambda <0$ & $\frac{3\lambda^2-\sqrt{48\lambda^2+9\lambda^4}}{48+9\lambda^2-3\sqrt{48\lambda^2+9\lambda^4}}$ & 1 & $\frac{-1+\frac{\lambda^2}{8}-\frac{\sqrt{48\lambda^2+9\lambda^4}}{24}}{1+\frac{3\lambda^2}{8}-\frac{\sqrt{48\lambda^2+9\lambda^4}}{8}}$ & $-1-\frac{1}{2}\lambda^2-\frac{\sqrt{48\lambda^2+9\lambda^4}}{6}$ & $\lambda <0$ \\
  \hline
\end{tabular}
\end{center}
\caption{\label{crit} The cosmological parameters and the behaviors of the critical points.}
\end{table*}

It is noteworthy that the stability of points do not mean the model is stable. So we must also investigate the stability of the model- both classical and quantum.
The classical fluctuations may be regarded to be stable when the speed of sound is positive, $c^{2}_{\rm s}\geq 0$. Combining this condition and Eq. (\ref{sound}), we
obtain the range in which the model is classic stable: $X\geq \frac{1}{2}$ ($x\geq 1$ or $x\leq -1$), or $X<\frac{1}{6}$ ($-\frac{\sqrt{3}}{3}<x<\frac{\sqrt{3}}{3}$).

In order to discuss the quantum stability of the scalar field, we consider the small fluctuations $\delta \phi$ around a background value $\phi_0$ which is a solution of the equations of motions: $\phi=\phi_0+\delta \phi$. Since we are most concerned with UV instabilities, it is not restrictive to consider a Minkowski background metric and to choose, at least locally, a time direction in such a way satisfy the decomposition. By expanding $p$ at the second order in $\delta \phi$, we can find the Lagrangian and the Hamiltonian fluctuations \cite{Piazza}:
\begin{eqnarray}
\label{fluc}
\delta \mathcal {H}=(p_X+2Xp_{XX})\frac{(\delta \dot{\phi})^2}{2}+p_X \frac{(\nabla \delta \phi)^2}{2}-p_{\phi\phi}\frac{(\delta \phi)^2}{2},
\end{eqnarray}
where $p_{\phi\phi}\equiv d^2p/d\phi^2$. The positivity of the first two terms in Eq. (\ref{fluc}) leads to the following stability conditions
\begin{eqnarray}
p_X+2Xp_{XX} \geq 0,~~~~~~p_X \geq 0.
\end{eqnarray}
Because $V(\phi)\geq 0$, we obtain $X\geq 1/2$ ($x\geq 1$ or $x\leq -1$) from these quantum stability conditions. So we conclude only for $X\geq 1/2$ ($x\geq 1$ or $x\leq -1$) the model is both classic and quantum stable. With these conditions we can determine whether the model is stable or not at critical points when variations $x$ and $y$ take the corresponding values of critical points.

In sum, the stability of points is related to the perturbations $\delta x$ and $\delta y$, and depends on the condition: Tr $\textbf{M}<0$ and $\det \textbf{M}>0$; the classic stability of the model is related to the perturbations $\delta \rho$, and depends on the condition: $c^{2}_{\rm s}\geq 0$; and the quantum stability of the model is related to the perturbations $\delta \phi$, and depends on the condition: $p_X+2Xp_{XX} \geq 0$ and $p_X \geq 0$. That is to say, the stability of the points is different from the classic/quantum stability of the model. So it is important to investigate not only the stability of points but also the stability of the model.

From Table \ref{crit}, it is obvious that points $P_1$, $P_2$, and $P_3$, are unstable for all $\lambda$; points $P_4$, ..., $P_9$ are stable for a certain range of $\lambda$. The model is classic stable at points $P_1$, ..., $P_7$ for all $\lambda$, classic unstable at points $P_8$ and $P_9$ for $\lambda \neq 0$; quantum stable at points $P_2$, ..., $P_7$ for all $\lambda$, quantum unstable at point $P_1$ for all $\lambda$, at points $P_8$ and $P_9$ for $0<\lambda<\sqrt{16/15}$.

\subsection{Cosmological implications}
According the phase-space analysis of a class of k-essence cosmology,
we can now discuss the corresponding cosmological behavior.
The critical points $P_{1}$, $P_{2}$, and $P_{3}$, corresponding to complete dark matter dominated universe,
are not relevant from a cosmological point of view, since they are unstable, though the model is classic and quantum stable at points $P_{2}$ and $P_{3}$ (quantum unstable at $P_{1}$).

Points $P_4$ with $\lambda<-\sqrt{3}$ and $P_5$ with $\lambda> \sqrt{3}$ are more interesting because not only the points are stable but also the model are both classic and quantum stable. Thus they can be the late-time attractor of the universe. In this case, k-essence behaves like dark matter; the final state seems like a dark matter dominated universe.

The critical points $P_6$ with $-\sqrt{3}<\lambda<0$ and $P_7$ with $0<\lambda<\sqrt{3}$ consist complete k-essence dominated and stable
late-time solutions, both presenting accelerating phases if $-\sqrt{3}/3<\lambda<0$ for $P_6$ and  $0<\lambda<\sqrt{3}/3$ for $P_7$. The model is
also both classic and quantum stable. For $P_6$ with $-\sqrt{3}<\lambda<-\sqrt{3}/3$ and $P_7$ with $\sqrt{3}/3<\lambda<\sqrt{3}$, these
two points present decelerating phases. For $P_6$ with $\lambda=-\sqrt{3}/3$ and $P_7$ with $\lambda=\sqrt{3}/3$, these
two points result to constant-speed expansion.

The critical points $P_8$ with $\lambda>0$ and $P_9$ with $\lambda<0$ also consist complete k-essence dominated and stable
late-time solutions, both presenting accelerating phases, at which k-essence behaves like phantom. But the model is classic unstable and quantum unstable ($0<\lambda<\sqrt{16/15}$) at these points. So they are also not physical interesting.

\section{Conclusions and discussions}
In this work we have made a comprehensive phase-space analysis of a class of
k-essence cosmology. We examined if a
universe governed by k-essence can have late-time
solutions compatible with observations.

We obtained the critical points and presented the conditions for their
existence and stability. We also calculated the values of the corresponding cosmological parameters, $c^{2}_{\rm s}$, $\Omega_{\phi}$, $w_{\phi}$, and $w_{\rm t}$, which are important in k-essence cosmology. We investigated the classic stability of the model, as well as the quantum stability. We discussed the behaviors of the critical points and plotted some of them. The critical points can be divided into three classes: unstable points at which the model is stable (for example, points $P_{2}$ and $P_{3}$), stable points at which the model is stable (for example, points $P_{4}$, $P_{5}$, $P_{6}$, and $P_{7}$), stable points at which the model is unstable (classic or quantum unstable, for example, points $P_{1}$, $P_{8}$, and $P_{9}$).
The case of points unstable but model stable or the case of points stable but model unstable are not relevant from a cosmological point of view. Only stable points at which the model is also stable (classic and quantum) are physically interesting. So only points $P_{4}$, $P_{5}$, $P_{6}$, and $P_{7}$ are cosmological relevant. For points $P_{4}$ and $P_{5}$, which can be the late-time attractor for the universe, at which k-essence behaves like dark matter, and the final state seems like a dark matter dominated universe. For points $P_{6}$ and $P_{7}$, which can also be the late-time attractor for the universe at which k-essence completely dominates, but all decelerating, constant-speed, and accelerating phases can appear, which dependents on the potential. In this case, we still can not determine what is the final state of the universe.
Although points $P_{8}$ and $P_{9}$ can indeed be the late-time attractor for the universe at which k-essence completely dominates and only accelerating phases appear, they are not physical concerning, because at which the model is both classic and quantum unstable.

The results we obtained have shown that the potential of k-essence is important for the behavior of the universe. In order to study the possible final state of the universe, it is important to investigate not only the stability of the critical points but also the (classic and quantum) stability of the model. To our knowledge, no previous work considers both of them at the same time. The case combining the stability of the points and non-adiabatic perturbation of the model is physical interesting and will be consider anywhere.

The analysis we performed indicates that a class of k-essence
cosmology discussed here can be compatible with observations. But it just faces the problem from the
cosmological point of view, and thus its results can been taken
into account only if k-essence passes successfully observational tests which is the subject of
interest of other studies.

\begin{acknowledgments}
This study is supported in part by Research Fund for Doctoral
Programs of Hebei University No. 2009-155, and by Open Research
Topics Fund of Key Laboratory of Particle Astrophysics, Institute of
High Energy Physics, Chinese Academy of Sciences, No.
0529410T41-200901.
\end{acknowledgments}

\bibliography{apssamp}

\end{document}